\documentclass{aa}  

\usepackage{graphics,graphicx}
\usepackage{xcolor}
\usepackage{enumitem}

\usepackage{txfonts}
\usepackage{longtable}
\usepackage[unicode=true,pdfusetitle,
 bookmarks=true,bookmarksnumbered=false,bookmarksopen=false,
 breaklinks=false,pdfborder={0 0 0},pdfborderstyle={},backref=false,colorlinks=true]
 {hyperref}
\hypersetup{
 urlcolor=blue,citecolor=blue}

\begin{document}

   \title{Probing the physical properties of the IGM using SRG/eROSITA spectra from blazars}
 
   \author{E. Gatuzz\inst{1}, 
           J. Wilms\inst{2}, 
           S. H\"ammerich\inst{2}, \and 
           R. Arcodia\inst{1,3} 
          }

   \institute{Max-Planck-Institut f\"ur extraterrestrische Physik, Gie{\ss}enbachstra{\ss}e 1, 85748 Garching, Germany\\
              \email{egatuzz@mpe.mpg.de}
         \and
             Dr. Karl Remeis-Observatory \& ECAP, Friedrich-Alexander-Universit\"at Erlangen-N\"urnberg, Sternwartstr. 7, 96049 Bamberg, Germany
         \and
             Kavli Institute for Astrophysics and Space Research, Massachusetts Institute of Technology, Cambridge, MA, USA
             }
 
   \date{Received XXX; accepted YYY}

  \abstract  
{Most baryonic matter resides in the intergalactic medium (IGM), a diffuse gas primarily composed of ionized hydrogen and helium, filling the space between galaxies. 
Observations of such an environment are crucial to better understanding the physical processes involved in such an environment. 
We present an analysis of the IGM absorption using blazar spectra from the first eROSITA all-sky survey (eRASS1) performed onboard of the Spectrum-Roentgen-Gamma mission (SRG) and {\it XMM-Newton} X-ray observations. 
First, we fitted the continuum spectra using a log-parabolic spectrum model and fixed the Galactic absorption. 
Then, we included a collisional ionization equilibrium model, namely {\tt IONeq}, to account for the IGM absorption. 
The column density $N({\rm H})$ and metallicity ($Z$) were set as free parameters. 
At the same time, the redshift of the absorber was fixed to half the blazar redshift as an approximation of the full line-of-sight absorber. 
We measured IGM-$N({\rm H})$ for 147 sources for SRG and 10 sources for {\it XMM-Newton}.  
We found a clear trend between IGM-$N({\rm H})$ and the blazar redshifts which scales as $(1+z)^{1.63\pm 0.12}$. 
The mean hydrogen density at $z=0$ is $n_{0}=(2.75\pm 0.63)\times 10^{-7}$ cm$^{-3}$. 
The mean temperature over the redshift range is $\log(T/K)=5.6\pm 0.6$ while the mean metallicity is  $Z=0.16\pm 0.09$. 
We found no acceptable fit using a power-law model for either temperatures or metallicities as a function of the redshift. 
These results indicate that the IGM contributes substantially to the total absorption seen in the blazar spectra.}

    \titlerunning{Studying the IGM with SRG/eROSITA blazars spectra}
    \authorrunning{Gatuzz et al.}
   \maketitle


\section{Introduction}\label{sec_in}      
The intergalactic medium (IGM) is a diffuse gas primarily composed of ionized hydrogen and helium, filling the space between galaxies within the vast cosmic structure known as the large-scale cosmic web. 
Most baryonic matter resides in the IGM, with a fraction even higher in the early universe as less material coalesced gravitationally from it \citep{mcq16}. 
Simulations predict that up to 50 percent of the baryons by mass have been shock-heated into the warm-hot phase (WHIM) at low redshift ($z<2$), with baryon densities $n_{b}=10^{-6}-10^{-4}$~cm$^{-3}$ and temperatures $T=10^{5}-10^{7}$~K \citep{cen99,dav07,sch15}. 
Simulations predict that the cool diffuse IGM constitutes $\sim 39\%$  of the baryons at redshift $z=0$ \citep{mar19}. 
In that sense, observations are essential to understand better the production, transport, and distribution of metals within the IGM. 

Most ionized IGM metals are not observed in optical to UV. Hence, X-ray absorption studies constitute a powerful technique to observe such an environment. 
Using a bright X-ray background source, acting as a lamp, the final X-ray absorption spectra provide information on the total absorbing column density of the material between the source and the observer. 
Typical sources commonly used to perform such analysis include active galactic nuclei (AGN) and gamma-ray bursts (GRBs) \citep{gal01,wat07,wat11,scha17,nic17,nic18,dal21a,gat23c}. 
After including the contribution of the Galactic absorption, which is usually known from 21~cm measurements \citep[e.g.,][]{kal05,wil13,hi416}, an additional absorption component is added to account for the ``excess absorption'' due to the IGM.  
With this technique, X-ray low-resolution spectra (i.e., CCDs) can be used to estimate the total hydrogen column density of the IGM. 
Future observatories, such as LEM \citep{kra22} and Athena \citep{nan13}, will measure X-ray high-resolution spectral features, such as lines and edges \citep{wal20}.

A key finding from previous observations of the IGM using high-redshift tracers is the apparent increase in excess of $N({\rm H})$ with redshift \citep[e.g.,][]{beh11,wat11,cam12}. 
One perspective postulates that the host source is responsible for all the excess and evolution \citep{owe98,gal01,str04,cam06,cam10,sch11,cam12,wat13,arc16,buc17}. Another perspective argues that the IGM contributes substantially to the absorption and, therefore, is redshift related \citep{sta13,cam15,rah19,dal20,dal21b,dal22}. Blazars, as sources with negligible X-ray absorption along the line of sight within the host galaxy, constitute ideal candidates for testing the absorption IGM component and the  $N({\rm H})-z$ relation. 
However, few studies have been done with such sources. 
\citet{arc18} analyzed {\it XMM-Newton} spectra of 15 blazars at $z>2$. 
They found an average IGM hydrogen density at redshift zero of $n_{0}=1.01^{+0.53}_{-0.72}\times 10^{-7}$ cm$^{-3}$ and temperature of $\log(T/K)=6.45_{-2.12}^{+0.51}$. 
\citet{dal21b}, on the other hand, analyzed a sample of 40 {\it Swift} blazars over a redshift range of $0.03<z<4.7$ in combination with a sub-sample of 7 {\it XMM-Newton} sources. 
They found that the $N({\rm H})$ scales as $(1+z)^{1.8\pm 0.2}$, with a mean hydrogen density at $z=0$ of $n_{0}=(3.2\pm 0.5)\times 10^{-7}$ cm$^{-3}$, a temperature of $\log(T/K)=6.1\pm 0.1$, and a mean metallicity of $[X/H]=-1.62\pm 0.04$ $(Z\sim 0.02)$.   
 
Here, we present an analysis of the IGM absorption using X-ray observations of blazars taken with the {\it SRG/eROSITA} X-ray telescope. 
This paper is structured as follows. 
Section~\ref{sec_dat} provides a detailed description of the data selection. 
Section~\ref{sec_fits} discusses the fitting procedure used in our analysis. 
We present the results in Section~\ref{sec_results}. 
We discuss and compare the results with other studies in Section~\ref{sec_dis}. 
Finally, Section~\ref{sec_con} summarizes our findings and presents our conclusions. 
Throughout this paper we assumed a $\Lambda$CDM cosmology with $\Omega_m = 0.3$, $\Omega_\Lambda = 0.7$, and $H_{0} = 70 \textrm{ km s}^{-1}\ \textrm{Mpc}^{-1} $.

\section{Data sample}\label{sec_dat} 
The {\it SRG/eROSITA} X-ray telescope \citep{pre21} on board of the {\it Spectrum Roentgen Gamma} (SRG) observatory consists of seven Wolter-I geometry X-ray telescopes providing all-sky survey observations in the $0.2-10$~keV energy range. 
During the first all-sky survey (eRASS1), more than 1 million sources were detected by {\it SRG/eROSITA}, with AGNs accounting for $80\%$ of the sources in the catalog (Merloni et al. 2023, in Review). 
An X-ray sample of blazars was created by {\bf H\"ammerich et al. (2023, submitted)} using eRASS1 data. 
Known blazars were taken from the 3HSP \citep{cha19}, ROMA BZCAT \citep{mas15} and 4FGL \citep{bal23} catalogs. 
By applying a cross-matching with a radius of 8\arcsec, 666 BL Lac and 841 flat spectrum radio quasars (FSRQ) were identified, the most significant systematic X-ray sample of blazars compiled so far. This dataset constitutes the initial sample in our analysis.

The complete details of the data reduction are shown in {\bf H\"ammerich et al. (2023, submitted)}, here we summarize the main aspects.
We reduced the data using the {\it SRG/eROSITA} data analysis software {\tt eSASS} pipeline version 020 \citep{bru22}.
First, we extracted the spectra using circular extraction regions with the radius scaled to the $0.2-2.3$~keV maximum likelihood (ML) count rate from the eRASS1 source catalog.
Following an empirical relation, larger radii correspond to higher count rates.
Background regions were created as annuli with size scaled to the ML count rate.
eRASS1 sources located within the background region were removed, using a circular region with radii depending on the ML count rate.
The task {\tt srctool} was used on the event files to create spectral and response files.
We combined the spectra from all Telescope Modules (TMs).

\subsection{XMM-Newton sub-sample}\label{sec_xmm_dat}
To cover $z>2$ redshift regions, we added a sub-sample of 15 blazars analyzed by \citet{arc18} in their study of the IGM absorption. Table~\ref{tab_obs} shows the sample details, covering a redshift range up to $z=4.715$. 
We reduced the spectra from the {\it XMM-Newton} European Photon Imaging Camera \citep[EPIC,][]{str01} with the Science Analysis System (SAS\footnote{https://www.cosmos.esa.int/web/xmm-newton/sas}, version 21.0.0).  
Observations were reduced with the {\tt epchain} SAS tool, including single and double pixel events (i.e., PATTERN$<=$4) and filtering the data with FLAG==0 to avoid bad pixels and parts of the detector close to the CCD edges.
Bad time intervals were filtered by using a 0.35 cts/s rate threshold.
For all sources, the final spectra were binned to oversample the instrumental resolution by at least a factor of 3 and to have a minimum of 20 counts per channel.  

\begin{table}[tb]
	\caption{List of $z>2$ with {\it XMM-Newton} sample.}
	\label{tab_obs}
	\centering
	\begin{tabular}{cccc} 
		\multicolumn{1}{c}{Name} &
		\multicolumn{1}{c}{RA} &
		\multicolumn{1}{c}{dec}&
		\multicolumn{1}{c}{$z$}  \\  
		\hline
		\\
		PKS 0528+134 	& 05 30 56.42 & +13 31 55.15 & 2.07\\ 
		4C 71.07  & 08 41 24.4 & +70 53 42 & 2.172\\
		QSO B0237-2322  & 02 40 08.18 & -23 09 15.78 & 2.225 \\
		PKS 2149-306 & 21 51 55.52 & -30 27 53.63 & 2.345 \\
		QSO J0555+3948  & 05 55 30.81 & +39 48 49.16 & 2.363 \\
		PBC J1656.2-3303 	& 16 56 16.78 & -33 02 12.7 & 2.4 \\
		QSO J2354-1513  & 23 54 30.20 & -15 13 11.16 & 2.675\\
		RBS 315  & 02 25 04.67 & +18 46 48.77& 2.69	\\
		QSO B0438-43  & 04 40 17.17 & -43 33 08.62& 2.852 \\
		QSO B0537-286  & 05 39 54.28 & -28 39 55.90& 3.104\\
		PKS 2126-158  & 21 29 12.18 & -15 38 41.02 & 3.268\\
		QSO B0014+810  & 00 17 08.48 & +81 35 08.14& 3.366 \\
		QSO B1026-084 & 10 28 38.79 & -08 44 38.44& 4.276  \\ 
		QSO J0525-3343  & 05 25 06.2 & -33 43 05 & 4.413 \\
		7C 1428+4218  & 14 30 23.74 & +42 04 36.49 & 4.715\\ 
		\hline
	\end{tabular}
\end{table}

\section{Spectral fitting}\label{sec_fits} 
All spectra were fitted using the {\it xspec} spectral fitting package (version 12.13.1\footnote{\url{https://heasarc.gsfc.nasa.gov/xanadu/xspec/}}). 
We use the {\tt cash} statistics \citep{cas79}. 
Errors are quoted at 1$\sigma$ confidence level unless otherwise stated. 
Finally, the abundances are given relative to \citet{gre98}.
We fitted the spectra in the $0.3-10$~keV energy range.
However, in the case of {\it SRG/eROSITA}, given that TM 5 and 7 are susceptible to excess soft emission due to optical light leaking \citep{pre21}, we created two groups by combining TM 1-4,6 and TM 5,7. 
The last one was analyzed only for energies $>1$~keV.
Below, we describe the spectral modeling.

\subsection{Galactic absorption + Continuum model}\label{sec_abs_mod}
Following the detailed study by \citet{dal21a}, we fitted each source with the model {\tt tbabs*ioneq*logpar}. 
The {\tt tbabs} component models the absorption in the local interstellar medium (ISM) as described by \citet{wil00}. 
The ISM-$N({\rm H})$ parameter was fixed to the \citet{wil13} values, which are estimated from 21~cm radio emission maps from \citet{kal05} but including a molecular hydrogen column density component. 
For the continuum, we included a log-parabolic spectrum {\tt logpar}, which can be produced by a log-parabolic distribution of relativistic particles \citep{pag09a} or due to a power-law particle distribution with a cooled high-energy tail \citep{fur13}. 
While a broken power-law intrinsic curvature could also arise from these environments, \citet{arc18} and \citet{dal21a} found that a log-parabolic lead to better modeling of blazar X-ray spectra, in good agreement with previous studies \citep[e.g., ][]{bha18,sah20}.

\subsection{IGM absorption}\label{sec_absigm_mod}
\begin{figure}    
\centering
\includegraphics[width=0.48\textwidth]{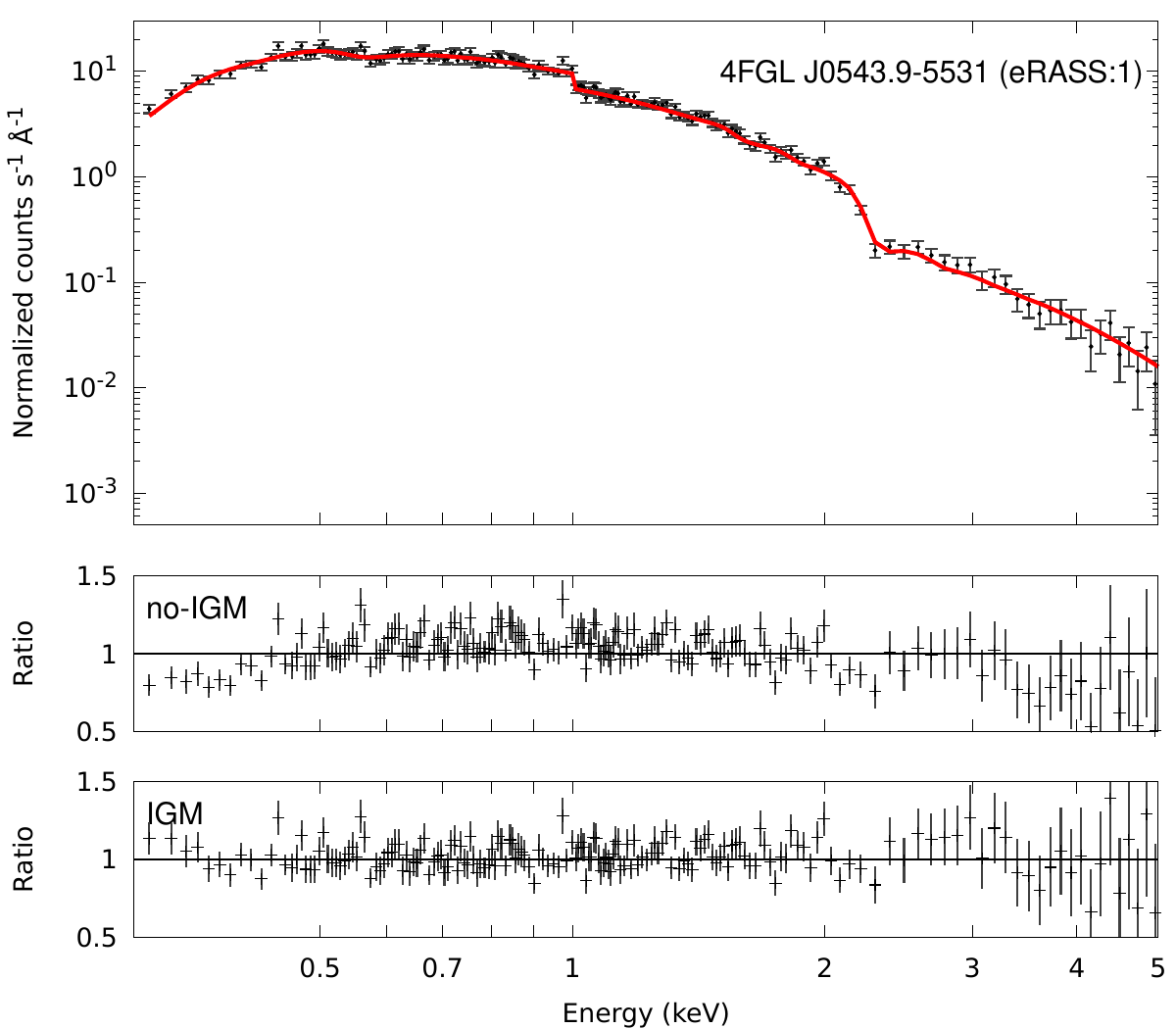}   
\caption{
{\it SRG/eROSITA} best-fit spectra of one of the sources included in the sample (4FGL J0543.9-5531). 
Black data points are the observations, while the solid red line correspond to the best-fit model including the IGM component. 
Bottom panels show the data-model ratio for the models without and with the IGM component (see Section~\ref{sec_fits} for further details). 
All cameras were grouped for illustration purposes.
}\label{fig_fit_example1} 
\end{figure}

\begin{figure}    
\centering
\includegraphics[width=0.48\textwidth]{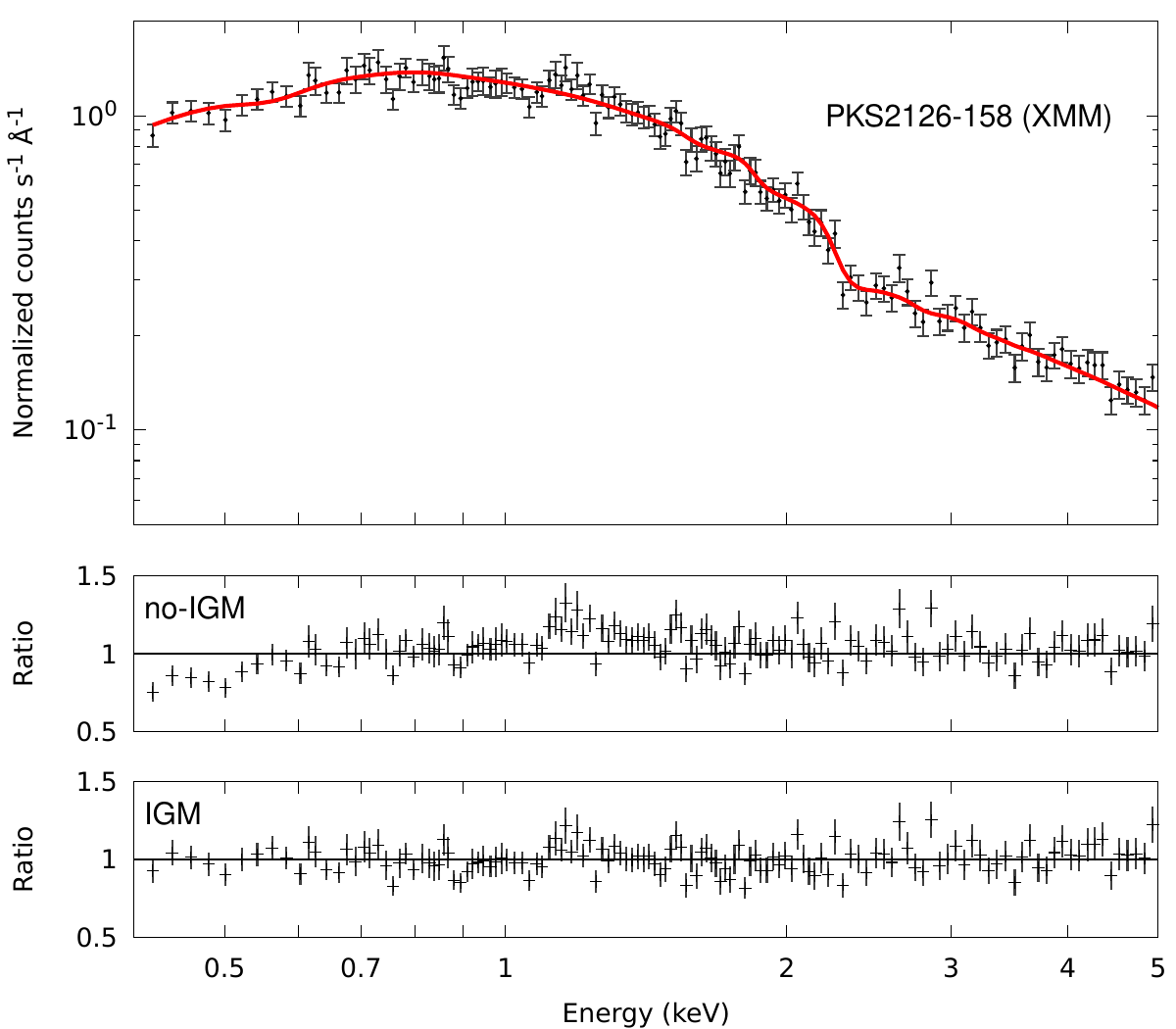}   
\caption{
{\it XMM-Newton} EPIC-pn best-fit spectra of one of the sources included in the sample (PKS2126-158, see Section~\ref{sec_fits} for further details).  
} \label{fig_fit_example2} 
\end{figure}

To model the IGM absorption we use the {\tt IONeq} model \citep{gat18} which assumes collisional ionization equilibrium (CIE) and includes as parameters of the model the hydrogen column density ($N({\rm H})$), gas temperature ($\log(T/K)$), metallicity ($Z$), turbulent broadening ($v_{turb}$) and redshift ($z$). 
Thus, we assume a thin uniform plane-parallel slab geometry in ionization equilibrium for the IGM. 
Such approximation is commonly used for a homogeneous medium \citep{sav14,kha19,leh19,dal21a}. 
We place this slab at half the blazar redshift to approximate the complete line-of-sight medium. 
The parameter ranges applied to this model are taken from \citet{dal21a} and listed in Table~\ref{tab_par}. 
The range of temperatures allows us to consider the cool IGM phase \citep[$4.5>\log(T/K)<5$,][]{sch03,sim04,agu08} and the warm-hot phases \citep[$\log(T/K)>5$,][]{dan16,pra18}. 
We note that the host X-ray absorption contribution in blazars is also negligible, swept by the kpc-scale relativistic jet \citep{arc18,dal21b}.
This is consistent with low optical-UV extinction levels observed in such sources \citep{pal16}.  
Simulations show that the density is very low along the line of sight to collimated outflows \citep{cha19a}. 
Figures~\ref{fig_fit_example1} and~\ref{fig_fit_example2} show examples of the best-fit results obtained for one source of the {\it SRG/eROSITA} and {\it XMM-Newton} samples.
The plots are zoomed into the soft-energy band to better illustrate the impact of adding an absorption component associated with the IGM. 
Finally, while a hybrid model including an absorber under photoionization equilibrium (PIE) conditions in combination with a CIE absorber would be more physical, high-resolution spectra would be necessary to test it \citep{dal21b}.

  \begin{table} 
\footnotesize
\caption{\label{tab_par}Upper and lower limits of the free parameters considered for the IGM absorption component.}
\centering 
\begin{tabular}{cccccccc}   
\hline
IGM-parameter& Range of values\\
\hline
Column density&$19\leq \log{(N({\rm H}))}\leq 23$\\    
Temperature&$4\leq \log{(T/K)} \leq 8 $\\ 
Metallicity&$-4\leq \log(Z) \leq -0.3 $\\  
 \hline
\end{tabular}
\end{table}

 \section{Spectral analysis results}\label{sec_results}
We have measured IGM-$N({\rm H})$ for 147 sources from the {\it SRG/eROSITA} initial sample, covering a redshift range up to $z=2.232$.
That is, the best-fit statistic was better when including the IGM component than those with only Galactic absorption \citep[$\Delta$c-stat $>10$, as in][]{dal21b}. 
Upper limits were obtained for 199 sources while for the rest of sources a fit statistical improvement was not found (i.e. 320 sources).
For the {\it XMM-Newton} sample we have measured IGM-$N({\rm H})$ for 10 sources and only upper limits for 5. 
In the following analysis, we have excluded the upper limits, which may lead to a biased understanding of the proper distribution of absorption features. 
In that sense, by incorporating a parameter range into our models, we ensure a comprehensive understanding of the uncertainties associated with our findings.

\subsection{IGM parameter results}\label{sec_results_igm}

\begin{figure}    
\centering
\includegraphics[width=0.5\textwidth]{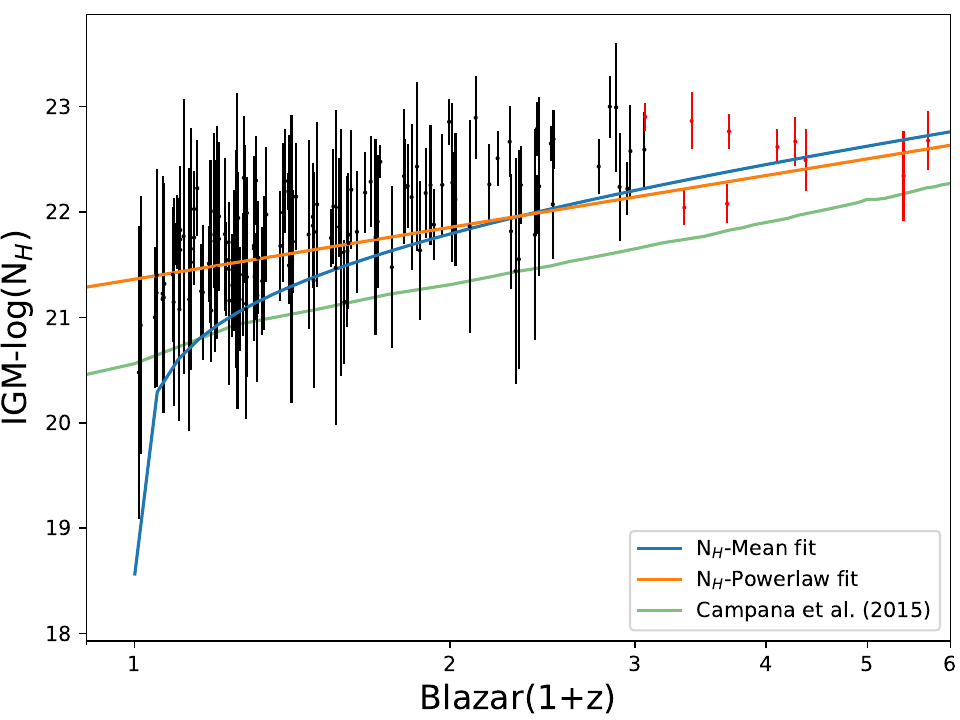}   
\caption{
Distribution of IGM-$N({\rm H})$ versus redshift. 
Black points belong to the {\it SRG/eROSITA} sample while red points correspond to the {\it XMM-newton} sample.
The blue line is the IGM model shown in Equation~\ref{eq_nh}. 
The solid orange line corresponds to a power-law fit.
The solid green line corresponds to the model proposed by \citet{cam15}.
} \label{fig_nhigm} 
\end{figure}

Figure~\ref{fig_nhigm} shows the IGM-$N({\rm H})$ distribution as a function of the redshift obtained from the best-fits. 
Black points correspond to the {\it SRG/eROSITA} data, while red points correspond to the {\it XMM-Newton} sample. 
The solid orange line in Figure~\ref{fig_nhigm} corresponds to a power-law fit of the IGM-$N({\rm H})$ versus the redshift trend for the complete sample.
It scales as $(1+z)^{1.63\pm 0.12}$ when including {\it XMM-newton} data and $(1+z)^{2.40\pm 0.19}$ when fitting only {\it SRG/eROSITA} data.
Such a fit is not appropriate for low redshift values, as can be seen from the plot. 
The solid blue line Figure~\ref{fig_nhigm} is the mean hydrogen density of the IGM based on the model developed in \citep[see ][and references therein]{sta13,shu18,dal21a}:

\begin{equation}\label{eq_nh}
N_{HXIGM}=\frac{n_{0}c}{H_{0}}\int_{0}^{z}\frac{(1+z)^{2}dz}{[\Omega_{M}(1+z)^{3}+\Omega_{\lambda}]^{\frac{1}{2}}}
\end{equation}
where $n_{0}$ is the hydrogen density at redshift zero, assuming that $90\%$ of the baryons are in the IGM \citep{beh11}. 
Using Equation~\ref{eq_nh}, we found a mean hydrogen density of $n_{0}=(2.75\pm 0.63)\times 10^{-7}$~cm$^{-3}$ when including the {\it XMM-Newton} while fitting only the {\it SRG/eROSITA} data we found $n_{0}=(4.31\pm 1.20)\times 10^{-7}$ cm$^{-3}$.   
This points out the importance of including sources at large redshift when doing IGM evolution studies.
 For comparison, Table~\ref{tab_no} shows a list of $n_{0}$ values obtained in previous works, including theoretical as well as IGM absorption measurements using blazars, quasars, and gamma-ray bursts (GRBs) as background sources. 
 The solid green line in Figure~\ref{fig_nhigm} corresponds to the model proposed by \citet{cam15}.
This curve is obtained from the cosmological simulations presented in \citet{pal13}.
At low redshift, we note that multiple blazars have larger IGM-$N({\rm H})$ than the mean hydrogen density model.
This could be due to the presence of the circumgalactic medium (CGM) in both our Galaxy and the host galaxy, or discrete intervening systems associated to other galaxies, thus providing more absorbing material.

  \begin{table} 
\footnotesize
\caption{\label{tab_no}$n_{0}$ values in literature.}
\centering 
\begin{tabular}{cccccccc}   
\hline
Reference& Value\\
\hline
This work& $(2.75\pm 0.63)\times 10^{-7}$ cm$^{-3}$\\    
\citet{beh11} - Theory& $1.7\times 10^{-7}$ cm$^{-3}$\\  
\citet{arc18} - Blazars&$(1.0\pm 0.6)\times 10^{-7}$ cm$^{-3}$\\  
\citet{dal21b} - Blazars&$(3.2\pm 0.5)\times 10^{-7}$ cm$^{-3}$\\ 
\citet{dal21a} - GRB&$(2.8\pm 0.3)\times 10^{-7}$ cm$^{-3}$\\   
\citet{dal22} - Quasars&$(2.8\pm 0.3)\times 10^{-7}$ cm$^{-3}$\\  
 \hline
\end{tabular}
\end{table}

\begin{figure}    
\centering
\includegraphics[width=0.5\textwidth]{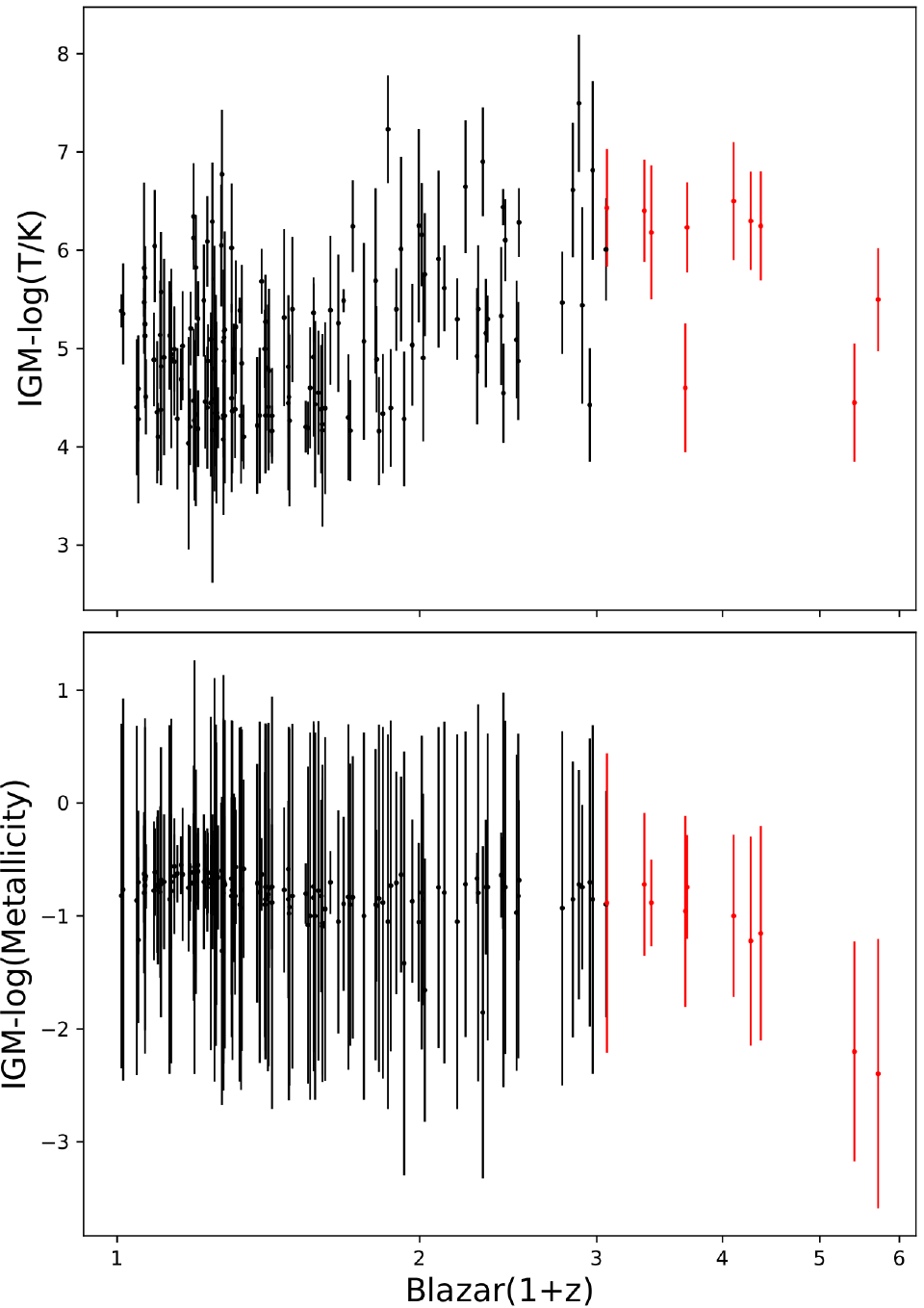}  
\caption{
\emph{Top panel:} Distribution of IGM-$kT$  versus redshift. 
\emph{Bottom panel:} Distribution of IGM-Metallicity  versus redshift. 
Black points belong to the {\it SRG/eROSITA} sample while red points correspond to the {\it XMM-newton} sample.
} \label{fig_kt_Z} 
\end{figure}

Top panel in Figure~\ref{fig_kt_Z} shows the temperature distribution as a function of the blazars redshift. 
Black points correspond to the {\it SRG/eROSITA} data, while red points correspond to the {\it XMM-Newton} sample.
The fitted temperatures scatter in the $4<\log(T/K)<7.5$ range.
The mean temperature over the redshift range is $\log(T/K)=5.6\pm 0.6$, consistent with the cold-WHIM \citep{tuo21}.
There is no apparent relation between the temperature and redshift due to the large scatter.
It is important to note because the fits are for the integrated line-of-sight, they are not representative of individual absorber temperatures.
Bottom panel in Figure~\ref{fig_kt_Z} shows the metallicity distribution as a function of the blazars redshift.
The fitted metallicities scatter in the $0.1 < Z < 0.2$ range at redshift $z<3$ with significant uncertainties.
The mean metallicity over the entire redshift range is $Z=0.16\pm 0.09$, which is consistent with the WHIM component \citep[e.g.,][]{nic18}.
While the metallicity tends to decrease with the redshift, we did not find an acceptable fit (i.e. $\chi^{2}/d.o.f.>5$) using a power-law model (i.e., $(1+z)^{a}$) for either temperatures or abundances.

 \subsection{Robustness of IGM-$N({\rm H})$ measurements}\label{add_test}
\begin{figure}    
\centering
\includegraphics[width=0.5\textwidth]{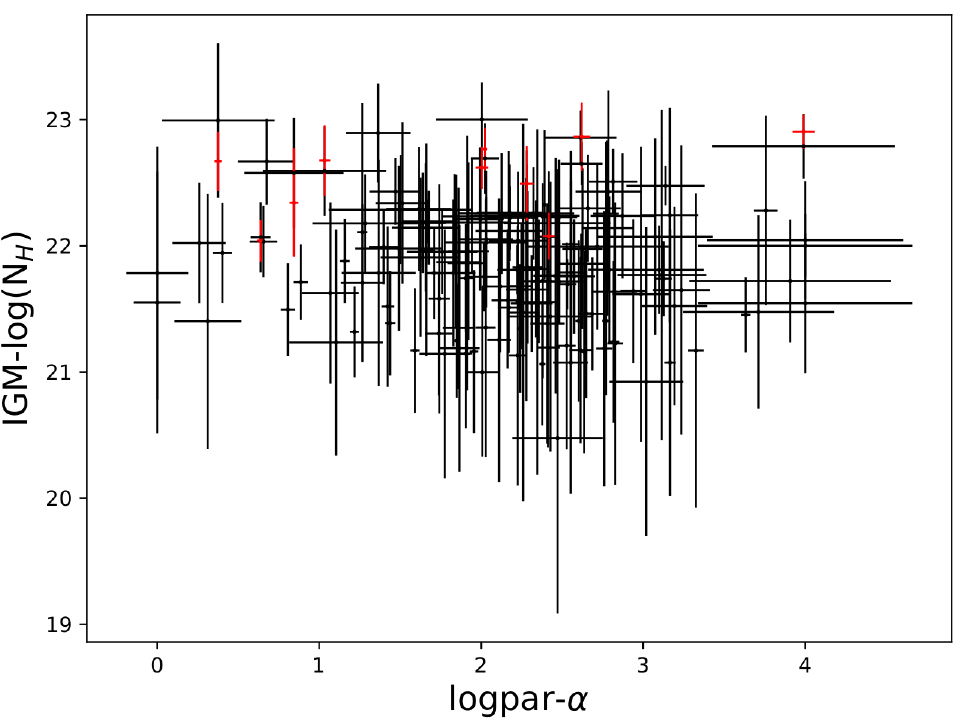}  \\ 
\includegraphics[width=0.5\textwidth]{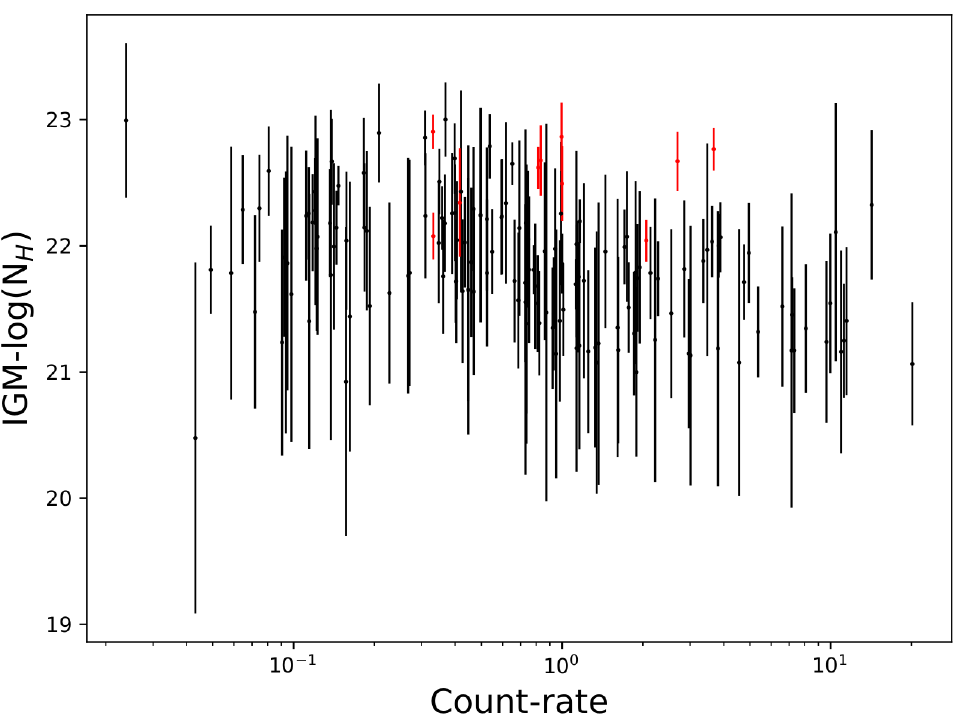} 
\caption{
\emph{Top panel:} Distribution of IGM-$N({\rm H})$ as function of the {\tt logpar} $\alpha$ indices. 
\emph{Bottom panel:} Distribution of IGM-$N({\rm H})$ as function of the count-rates. 
Black points belong to the {\it SRG/eROSITA} sample while red points correspond to the {\it XMM-newton} sample.
} \label{fig_nh_alpha_rate} 
\end{figure}

Various scenarios could account for the observed soft excess in blazars without necessitating IGM absorption. 
For example, an IGM-$N({\rm H})$ and spectral degeneracy could happen to some degree.  
A harder spectrum slope may mimic large IGM-$N({\rm H})$ and vice versa.
We checked for relation between the IGM-$N({\rm H})$ and the $\alpha$ index of the {\tt logpar} model (see top panel in Figure~\ref{fig_nh_alpha_rate})
We found no apparent relation between $\alpha$ indices and the column density.
This is consistent with previous results \citep{arc18,hai19,dal21a}
The bottom panel in Figure~\ref{fig_nh_alpha_rate} shows the count rate for each source as a function of the IGM-$N({\rm H})$.
There is no apparent relation between both parameters. 
Blazars display variability due to multiple processes such as particle acceleration, injection, cooling, and escape \citep{gau20}.
Any IGM absorber found in the LOS to such a system should not show variability, while a local absorber can.
While eRASS1 data alone cannot be used to study variability within the sources, \citet{dal21a} shown in their analysis of {\it XMM-newton} and {\it Swift} blazar spectra that the lack of apparent relation between  IGM-$N({\rm H})$ and flux indicates that the absorption measured from the X-ray spectra is associated with the IGM rather than the sources. 
Finally, it is notable that our model relies on the assumption of a thin slab positioned at half the redshift of each source.
The efficacy of this approach is contingent upon the validity of this simplifying assumption and may be subject to the specific characteristics of the observed blazar spectra.
A thorough examination of this approximation across various sources and conditions with future X-ray observatories such as {\it Athena} \citep{nan13} would provide valuable insights into its robustness.

  \section{Comparison with previous works}\label{sec_dis}
Despite their ideal properties to study such an environment, only a few systematic IGM absorption studies have been done using blazars as X-ray sources. 
\citet{arc18} measured IGM-$N({\rm H})$ for a sample of 15 sources using {\it XMM-newton} observations by using an {\tt absori} based {\sc xspec} absorption model. 
They concluded that excess absorption is the preferred scenario to explain the $N({\rm H})-(z+1)$ relation. 
They found an IGM gas density of $(1.0\pm 0.6)\times 10^{-7}$ cm$^{-3}$ and a temperature of $\log(T/K)=6.45^{+0.53}_{-0.72}$. 
When comparing with their results, apart from differences in both the continuum and the absorption modeling, we noted that they fixed the continuum {\tt powerlaw} parameters when including the IGM absorber, an approach adopted due to computational limits. 
More importantly, they assumed a solar abundance in their modeling.  
From the IGM gas density they inferred an IGM metallicity $Z_{IGM}=0.59^{+0.31}_{-0.42}$, which is larger than the value obtained from our analysis. 
\citet{dal21b} analyzed {\it Swift} spectra of 40 blazars with a redshift range of $0.03\leq z \leq 4.7$. 
They also included {\it XMM-Newton} observations of 7 blazars in the $0.86\leq z \leq 3.26$ range. 
We follow the same analysis methodology described in Section~\ref{sec_fits}. 
Considering uncertainties, our $n_{0}$ and metallicity values are in good agreement with their results (see Table~\ref{tab_no}). 
Moreover, they found no apparent relation between the {\tt logpar} index or the column density.

  \section{Conclusions}\label{sec_con}
We study the IGM absorption by using the X-ray sample of blazars computed by {\bf H\"ammerich et al. (2023, submitted)} using eRASS1 data in combination with a {\it XMM-newton} sub-sample of 15 blazars. 
Following the work done in \citet{dal21b}, we modeled the continuum with a log-parabolic spectrum model and fixed the Galactic absorption to the 21~cm values. 
After modeling the continuum component, we included a {\tt IONeq} model, which assumes collisional ionization equilibrium conditions, to search for IGM absorbers. 
The column density $N({\rm H})$ and metallicity ($Z$) were set as free parameters. At the same time, the redshift of the absorber was fixed to half the blazar redshift as an approximation of the full line-of-sight absorber. 
From the {\it SRG/eROSITA} sample, we measured column densities (IGM-$N({\rm H})$) for 147 sources, covering a redshift range up to $z=2.232$ while for the {\it XMM-Newton} sample we measured IGM-$N({\rm H})$ for 10 sources. 
We found a clear trend between IGM-$N({\rm H})$ and the blazar redshifts, which can be modeled with a mean hydrogen density model with $n_{0}=(2.75\pm 0.63)\times 10^{-7}$ cm$^{-3}$.  
The best-fitting model corresponds to $(1+z)^{1.63\pm 0.12}$.
We found no acceptable fit using a power-law model for either temperatures or metallicities as a function of the redshift. 
We conclude that the absorption measured is not source-related but due to the IGM, contributing substantially to the total absorption seen in the blazar spectra. 
Finally, this work will be followed by a detailed study of the IGM absorption using {\it SRG/eROSITA} quasars spectra.

\begin{acknowledgements} 
The authors thank Sergio Campana and Ruben Salvaterra for providing their model results in Figure~\ref{fig_nhigm}.
This work is based on data from {\it SRG/eROSITA}, the soft X-ray instrument aboard SRG, a joint Russian-German science mission supported by the Russian Space Agency (Roskosmos), in the interests of the Russian Academy of Sciences represented by its Space Research Institute (IKI), and the Deutsches Zentrum f\"ur Luft- und Raumfahrt (DLR). 
The SRG spacecraft was built by Lavochkin Association (NPOL) and its subcontractors, and is operated by NPOL with support from the Max Planck Institute for Extraterrestrial Physics (MPE). 
The development and construction of the {\it SRG/eROSITA} X-ray instrument was led by MPE, with contributions from the Dr. Karl Remeis Observatory Bamberg \& ECAP (FAU Erlangen-Nuernberg), the University of Hamburg Observatory, the Leibniz Institute for Astrophysics Potsdam (AIP), and the Institute for Astronomy and Astrophysics of the University of T\"ubingen, with the support of DLR and the Max Planck Society. 
The Argelander Institute for Astronomy of the University of Bonn and the Ludwig Maximilians Universit\"at Munich also participated in the science preparation for {\it SRG/eROSITA}. 
This work is based on observations obtained with {\it XMM-Newton}, an ESA science mission with instruments and contributions directly funded by ESA Member States and NASA. 
This research was carried out on the High Performance Computing resources of the cobra cluster at the Max Planck Computing and Data Facility (MPCDF) in Garching operated by the Max Planck Society (MPG). 
The {\it SRG/eROSITA} data shown here were processed using the eSASS software system developed by the German {\it SRG/eROSITA} consortium.   
\end{acknowledgements}

%
%
\bibliographystyle{aa}
 \newcommand{\noop}[1]{}

\end{document}